\documentclass[fleqn,usenatbib]{mnras}

\usepackage[T1]{fontenc}

\DeclareRobustCommand{\VAN}[3]{#2}
\let\VANthebibliography\thebibliography
\def\thebibliography{\DeclareRobustCommand{\VAN}[3]{##3}\VANthebibliography}

\usepackage{graphicx}	
\usepackage{amsmath}	
\usepackage{amssymb}	
\usepackage{newtxtext,newtxmath}
\usepackage[dvipsnames]{xcolor}
\usepackage{multirow}
\usepackage{xcolor}
\usepackage{{booktabs}}
\usepackage{{longtable}}
\usepackage{pdflscape}
\usepackage{caption}
\usepackage{subcaption}
\usepackage{booktabs}

\newcommand{\ergps}{erg\,s$^{-1}$}
\newcommand{\ergpspcm}{erg \,s$^{-1}$\,cm$^{-2}$}

\newcommand{\msun}{M_\odot}

\defcitealias{Quintin2023}{Q23}

\title[QPE in a TDE with eROSITA]{Further evidence of Quasiperiodic Eruptions in a tidal disruption event AT2019vcb by SRG/eROSITA}

\author[Bykov et al.]{
S.D. Bykov$^{1,2}$\thanks{E-mail:sergei.d.bykov@gmail.com}, M.R. Gilfanov$^{3,2}$, R.A. Sunyaev$^{3,2}$ and P.S.Medvedev$^{3}$\\
$^{1}$Kazan Federal University, 18 Kremlyovskaya Street, Kazan, Russia\\
$^{2}$Max Planck Institute for Astrophysics, Karl-Schwarzschild-Str 1, Garching b. München D-85741, Germany\\
$^{3}$Space Research Institute, Russian Academy of Sciences, Profsoyuznaya 84/32, 117997 Moscow, Russia\\
}

\date{Accepted XXX. Received YYY; in original form ZZZ}

\pubyear{2024}

\begin{document}
\label{firstpage}
\pagerange{\pageref{firstpage}--\pageref{lastpage}}
\maketitle

\begin{abstract} 
We report the discovery of a short, large amplitude X-ray flare from AT2019vcb (aka Tormund), a tidal disruption event (TDE) at $z=0.088$. The discovery is based on the data from the SRG/eROSITA X-ray telescope which happened to observe the source seven months after the onset of the optical TDE. eROSITA observation occurred 13 days after a soft flare was detected in the \textit{XMM-Newton} data. Both events bear similar characteristics in terms of timing and spectral properties. eROSITA spectrum is described as an accretion disk with a characteristic temperature of $\sim180$ eV and luminosity $\sim8\times10^{43}$ \ergps. The eROSITA flare lasted less than 12 hours and had an amplitude $\ge70$ with respect to the quiescent level, no flares were detected in later eROSITA observations (6-18 months later). The \textit{XMM-Newton} and eROSITA flares provide strong evidence that the TDE AT2019vcb is a bona fide QPE source. Our work further strengthens the direct connection between TDEs and QPE following similar recent results in a TDE AT2019qiz. 


\end{abstract}




\section{Introduction}


X-ray quasi-periodic eruptions (QPE) are dramatic transient events associated with massive black holes in galactic centres.  They are characterised by strong, repeating flares of soft X-ray emission superimposed to a stable thermal-like quiescent level, likely from an accretion flow. The observed periods/recurrence times are on the time-scale of 30 minutes to days, with outbursts lasting hours. The luminosities achieved are of the order of $10^{42}$--$10^{44}$ \ergps. The X-ray spectra are black body-like with typical temperatures of $\sim100-200$ eV. More complicated phenomenology has also been observed, such as energy-dependent light curves, evolution of temperature, QPE lifetime, recurrence time, and similar observable characteristics (\citealt{Miniutti2019, Chakraborty2021, Arcodia2022,  Chakraborty2024, Pasham2024, Giustini2024}).  

Only a handful of QPEs are known -- seven sources and two candidates. GSN 069 \citep[the first QPE discovered by][]{Miniutti2019}, RX J1301.9+2747 \citep{Giustini2020}, eRO-QPE1,2,3,4 \citep[discovered by SRG/eROSITA; ][]{Arcodia2021,Arcodia2024a}, AT2019qiz \citep[the most recenty discovered QPE by][]{Nicholl2024}. The two candidates are XMMSL1 J024916.6-041244 \citep[aka J0249; ][]{Chakraborty2021} and AT2019vcb/4XMM J123856.3+330957 \citep[aka Tormund, ][]{Quintin2023}.  In addition to these seven sources  we mention Swift J0230 \citep{Evans2023, Guolo2024} which represents a phenomenon similar to longer-period repeating TDEs, which might be related to QPEs but most likely not coincident with it. 
 In the present paper we study one of the two QPE candidates listed above, AT2019vcb/4XMM J123856.3+330957, 
for which we present further evidence that this is indeed a QPE source.

There is no definitive model explaining all observed properties of QPEs. The processes related to orbital motion are thought to be responsible for the phenomena, such as gravitational self-lensing \citep{Ingram2021} or binary black holes interaction and extreme mass ratio inspirals, EMRI, \citep{Shu2018, King2020, Xian2021, King2022, Metzger2022, Zhao2022, Lu2023, Franchini2023, Tagawa2023, Krolik2022, Linial2023, Zhou2024a, Zhou2024b},  Alternatively, accretion flow instabilities are considered but less preferred \citep{Raj2021, Pan2022, Krolik2022, Kaur2023}

It appears that QPEs occur in galaxies with central black holes of relatively small masses ($10^5-10^6 \msun$) and absent or weak AGN activity \citep{Wevers2022, Arcodia2024b}. GSN 069, Tormund and J0249 seem to be connected to a tidal disruption event (TDE) which had occurred before QPEs were detected.   Also, GSN 069 shows abnormal C/N ratio in both UV and X-rays \citep{Sheng2021, Kosec2025}, similarly to other TDEs \citep{Kochanek2016, Yang2017, Miller2023}. The X-ray decay of the quiescent emission in eRO-QPE3 is also suggestive of a TDE \citep{Arcodia2024a}.
 A TDE is a disruption of a star by a massive black hole, which may cause a huge, month-to-year-long, outburst of radiation \citep{Rees1988, Gezari2021}.  The similarity between host galaxies of TDEs and QPEs also hints at  a connection \citep{Wevers2022}. Moreover, recently, many QPE flares were discovered in a TDE AT2019qiz by \citet{Nicholl2024}, making an almost direct connection between the two populations.

In this paper, we study a transient object AT2019vcb (aka Tormund). AT2019vcb was discovered by the Zwicky Transient Facility (ZTF, \citealt{Bellm2019}) in 2019 and classified as a tidal disruption event \citep{Hammerstein2023}. The host has a position RA=$189.734917$, Dec=$+33.165911$ deg and redshift $z=0.088$. \citet[][hereafter Q23]{Quintin2023} present a detailed analysis of X-ray data from \textit{XMM-Newton}, \textit{Swift/XRT} and \textit{NICER} observatories following up this TDE.  The main finding of their work was the detection of a start of a bright flare which occurred at the end of the \textit{XMM-Newton} observation. The similarity of its temporal and spectral properties to other QPEs, in particular to eRO-QPE1 has led the authors to the conclusion that they may have detected the rising part of a QPE flare, rather than simple rebrightening of the TDE light curve (see figs 5, 7, 13 in \citetalias{Quintin2023}). Peak luminosity detected is $1.2\times10^{44}$ \ergps, with a rising time of around four hours. The flare occurred on May 22, 2020. The black body spectral temperature rose from 50 to 110 eV. Follow-up of the source with the \textit{NICER} telescope two years later shows that the source is active (possibly a TDE re-brightening), but no QPEs were observed. 

Here we report on the serendipitous detection of another soft, short-duration flare from this source by SRG/eROSITA telescope on June 3, 2020, merely 13 days after the \textit{XMM-Newton} flare. In sect. \ref{sect:data} we explain the eROSITA data collection and light curve aggregation. In sect. \ref{sect:analysis} we report our findings: flare light curves, spectral analysis and upper limits. We briefly discuss the implication of our finding in sect. \ref{sect:discussion}.  We conclude this report in sect. \ref{sect:conclusion}.  Errors are $1\sigma$ and upper limits are $90\%$ confidence unless stated otherwise.

\section{SRG/\lowercase{e}ROSITA Data}
\label{sect:data}


We utilise the data of the eROSITA telescope \citep{Predehl2021} aboard SRG satellite  \citep{Sunyaev2021} collected during the all-sky survey in 2019-2022. The all-sky survey started in December 2019, in February 2022, the eROSITA telescope was switched to the safe mode and stopped registering X-ray photons, while the SRG observatory continued observations in the interests of the ART-XC telescope. The sky is fully scanned in 6 months and during 2019-2022 eROSITA telescope surveyed the full sky four times and about 38\% of the sky -- five times.  The eROSITA X-ray telescope collects data in the 0.2--9 keV energy range. Its detailed description can be found in \citet{Predehl2021}.  eROSITA data calibration and reduction was performed with the eSASS software developed by the German eROSITA consortium \citep{Brunner2022} and the software produced by the Russian SRG/eROSITA consortium. 
X-ray spectral fitting was performed in the $0.2$-$9.0$ keV energy range using Cash statistic \citep{Cash1979}. 
We checked that using $0.3-9.0$ keV energy range for telescope modules 5 and 7 affected by optical light leakage \citep{Predehl2021} does not alter our results. The light curves and spectra are obtained via forced aperture photometry unless stated otherwise.  Errors on the count rate from light curves are obtained using formulas from \citet{Gehrels1986}. 

Given the satellite's scanning strategy, during each all-sky survey (eRASS), a source is typically visited six times for 40 seconds per visit, with a cadence of 4 hours \citep{Sunyaev2021}. The number of 40-second visits (passes) depends significantly on the ecliptic latitude. We ignore passes with large angles between the telescope's axis and the source using condition \texttt{FRACEXP}$>0.05$.  The object studied here, AT2019vcb,  was typically scanned 7-8 times in each of eRASS1-4 (29 passes total). No eRASS5 data is available.  

\section{\lowercase{e}ROSITA analysis}
\label{sect:analysis}

\subsection{Light curve}

There is a bright source in the SRG/eROSITA source catalogue,  SRGe J123856.5+330954, coincident with the position of AT2019vcb. The eROSITA source is located at RA=$189.735603$, Dec$=+33.164982$ deg, in $3.9"$ from AT2019vcb. Its positional uncertainty is $5.6"$ (98\% confidence), making association with AT2019vcb unambiguous. It has a detection likelihood $\mathrm{DL}=216.0$ in the $0.3-2.3$ keV energy range, which is far above the detection threshold for the catalogue construction ($\mathrm{DL}=6.0$). The PSF photometry yielded  83 photons from the source in the combined data of eRASS1-4.
Averaged over all-sky surveys, the source has a count rate of $0.16\pm0.02$ counts/s, which roughly corresponds to the X-ray flux of $\sim10^{-13}$ \ergpspcm.

\begin{figure*}
    \centering
    \includegraphics[width=1\textwidth]{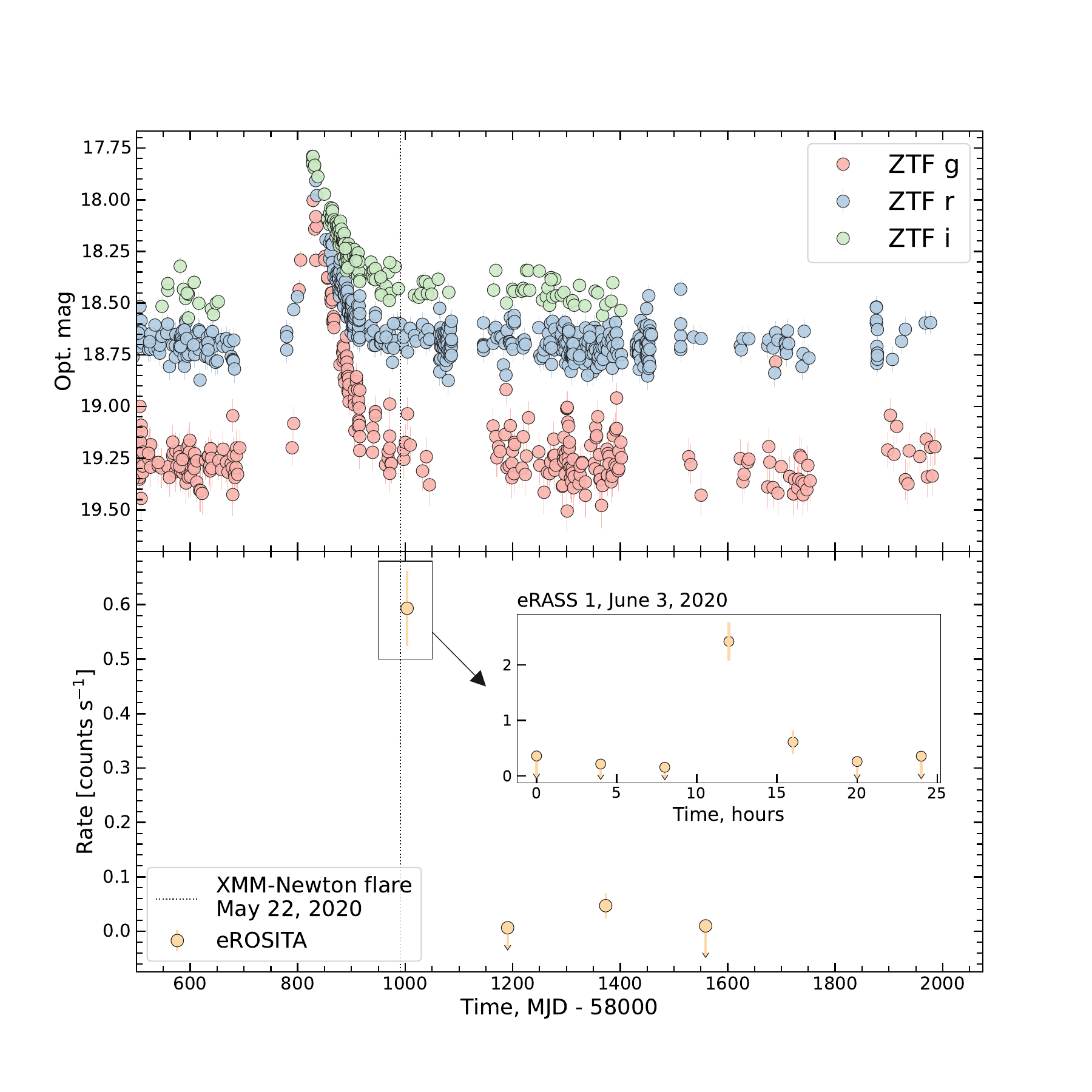}
    \caption{Light curve of AT2019vcb. The top panel shows the optical (ZTF) light curve of the tidal disruption event. The bottom panel shows the eROSITA X-ray light curve, with the inset showing short-term variability detected in eRASS1. Both panels have a mark at the date of the \textit{XMM-Netwon} flare detected by \citeauthor{Quintin2023}.}
    \label{fig:lc}
\end{figure*}

SRGe J123856.5+330954 is strongly variable. From PSF photometry,  in eRASS1, it averaged a count rate of $0.6\pm0.06$ counts/s, in eRASS3 -- $0.04\pm0.02$ counts/s (detection likelihood $\mathrm{DL}=6.6$, which slightly exceeds $3\sigma$), and in eRASS 2,4 the source is below the detection limit. The X-ray light curve is shown in the bottom panel in Fig. \ref{fig:lc}. The upper panel displays the optical (ZTF) light curve showing the TDE rise and fall and marks a date when \citetalias{Quintin2023} detected a QPE-like feature.  The ZTF light curves were retrieved from the \href{http://db.ztf.snad.space}{SNAD service} \citep{Malanchev2023}.

The detailed light curve of eRASS1 (inset in Fig \ref{fig:lc}) reveals strong variability on the 4-hour time-scale.  The first three passes show no sign of X-ray emission. In the fourth scan, the source becomes very bright, reaching a flux of around 2.5 counts/s. After 4 hours, the source is still significant with 0.6 counts/s. After another 4 hours, the source is on the detection limit with 0.1 counts/s. No signs of short variability are visible in later data (in eRASS2-4) --  the eROSITA light curves for all scans are shown in Fig. \ref{fig:lc_all}. The fluxes/upper limits and luminosities are calculated in the next subsection.  The X-ray luminosity evolution is presented in Fig. \ref{fig:lc_lumin}.

\subsection{Spectra}

The amount of photons detected by eROSITA allows meaningful spectral analysis. We select several time periods for measuring spectra (see Table \ref{tab:erosita}) via forced aperture photometry. The first period (A) encompasses the first three passes in eRASS1. Segments B and C are the 4th and 5th passes (when the source is bright) respectively. Period D are the last two scans of eRASS 1. Segments E, F, and G are the eRASS2,3,4 respectively.

Spectral modelling is performed assuming the spectral model of multi-colour accretion disc \texttt{diskbb} (redshifted to $z=0.088$ with \texttt{zashift}), and with the interstellar absorption fixed at the value reported by \citet{HI4PICollaboration2016} at the position of the source: $N_{\rm H}=1.45~\times~10^{20}$ cm$^{-2}$. The \textsc{xspec} model is therefore \texttt{tbabs*zashift*diskbb}. Unabsorbed fluxes are measured with the \texttt{cflux} component (\texttt{tbabs*zashift*cflux*diskbb}). Results of the redshifted black body model \texttt{tbabs*zashift*bbodyrad} are also reported in the appendix (Table \ref{tab:erosita_bbodyrad}). 

Table \ref{tab:erosita} shows the spectral modelling results. The temperature is reliably measured only in period B, all other segments have temperature was frozen at that value. Statistically significant flux is measured only in periods B and C, for other intervals we obtained upper limits on the flux/luminosity. We describe the results for the accretion disc model, but the conclusions for the black body radiation spectrum are similar albeit with a slightly lower best-fitting temperature. 

\begin{figure}
    \centering
    \includegraphics[width=0.5\textwidth]{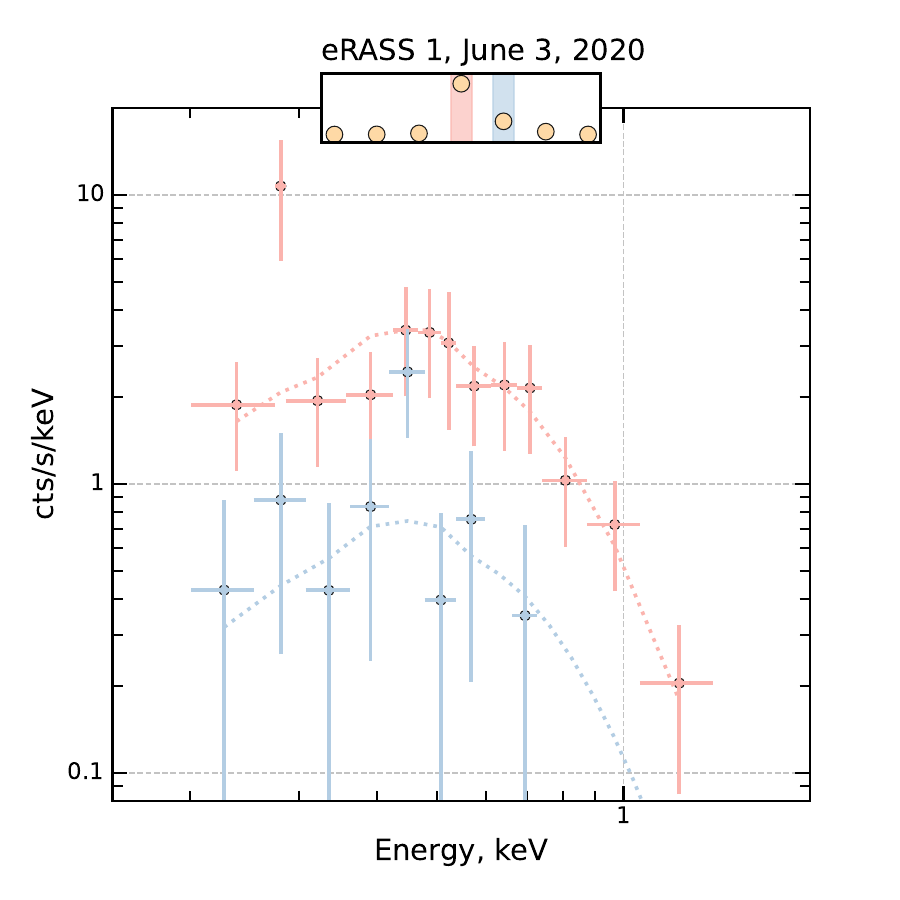}
    \caption{eROSITA X-ray spectra for segments B and C (presumed QPE flare). The spectra are \texttt{data} mode for \textsc{xspec} plotting (circles with error bars). Lines show the best-fitting \texttt{diskbb} models. The inset on the top shows the 4-hour light curve of eRASS 1, on June 3, 2020, with the corresponding colour of a segment highlighted. Data is rebinned for clarity.}
    \label{fig:spe}
\end{figure}

Segment B is fitted with the \texttt{diskbb} (\texttt{bbodyrad}) model of temperature $\approx170$ eV ($\approx130$ eV) and characteristic radius of $\approx0.7\times10^6$ km ($\approx2\times10^6$ km) assuming face-on orientation of the accretion disk (black body). The fit statistic is acceptable, with the disc model being slightly more preferable compared to the black body (c-stat 20.7 vs 24.4 respectively for 21 dof). Luminosity in segment B is around $8\times10^{43}$ \ergps ($7\times10^{43}$ \ergps) in the 0.2--2.0 keV range. For the next segment (C), we could not constrain the normalisation and temperature simultaneously, hence we fixed the temperature at the value from segment B. This results in the emission region size of $\approx0.5\times10^6$ km ($\approx10^6$ km) respectively, with the luminosity $\approx 3\times10^{43}$ \ergps. Fig. \ref{fig:spe} shows the observed spectra for segments B and C, along with their best-fitting models. 

All other segments were fitted with the fixed temperature model and only upper limits for flux/luminosity could be obtained. Upper limits are in the $10^{41}-10^{42}$ \ergps\,  range \footnote{The forced PSF photometry analysis gave a weak detection of the source in eRASS3, slightly over $3\sigma$ level (see Fig. \ref{fig:lc}). However, as it often happens, the procedure used for flux/luminosity estimation via \textsc{xspec} fitting did not result in a statistically significant flux measurement, with the upper limit larger than the flux obtained from PSF fitting. We therefore conservatively assigned to eRASS3 the upper limit obtained in \textsc{xspec}.}. The flare amplitude (compared with the pre-flare upper limit) is therefore at least 70.

The overall picture is that the source is virtually undetected in all eROSITA passes except two. During the two consecutive passes, the source is bright and soft.

{\fontsize{7}{9}\selectfont
\begin{table*}
\centering
\begin{tabular}{lllllll}
\toprule
eRASS$^{(1)}$  &  Period$^{(2)}$  &                        Date$^{(3)}$  &  $F_{\rm X}$, erg s$^{-1}$ cm$^{-2}$ $^{(4)}$  &    $L_{\rm X}$, erg s$^{-1}$$^{(5)}$  &  T$_{\rm in}$, eV$^{(6)}$  &  $R_{\rm in}$/$\cos\theta$, km$^{(7)}$ \\
\midrule
      eRASS 1  &               A  &             2020-06-03 02:22--10:22  &                         $<6.0 \times 10^{-14}$ &                 $<1.1 \times 10^{42}$ &                    $173^*$ &                                      - \\
      eRASS 1  &               B  &                    2020-06-03 14:22  &         $4.4 \,^{+0.5}_{-0.6} \times 10^{-12}$ & $7.9 \,^{+0.9}_{-1.0} \times 10^{43}$ &       $174 \,^{+15}_{-16}$ &         $7 \,^{+5}_{-6} \times 10^{5}$ \\
      eRASS 1  &               C  &                    2020-06-03 18:22  &         $1.8 \,^{+0.4}_{-0.5} \times 10^{-12}$ & $3.1 \,^{+0.8}_{-0.9} \times 10^{43}$ &                    $173^*$ &   $4.4 \,^{+2.1}_{-2.3} \times 10^{5}$ \\
      eRASS 1  &               D  &  2020-06-03 22:22--2020-06-04 02:22  &                         $<1.7 \times 10^{-13}$ &                 $<3.0 \times 10^{42}$ &                    $173^*$ &                                      - \\
      eRASS 2  &               E  &  2020-12-07 01:07--2020-12-08 01:07  &                         $<3.4 \times 10^{-14}$ &                 $<6.2 \times 10^{41}$ &                    $173^*$ &                                      - \\
      eRASS 3  &               F  & 2021-06-07 11:22--2021-06-08 15:22   &                         $<1.0 \times 10^{-13}$ &                 $<1.9 \times 10^{42}$ &                    $173^*$ &                                      - \\
      eRASS 4  &               G  & 2021-12-10 14:07--2021-12-11 14:07   &                         $<3.1 \times 10^{-14}$ &                 $<5.6 \times 10^{41}$ &                    $173^*$ &                                      - \\
\bottomrule

\end{tabular}

 \caption{SRG/eROSITA observations of AT2019vcb. (1) - all-sky sky number; (2, 3) particular observation date;  (4,5) X-ray flux and luminosity in the 0.2--2.0 keV energy range; (6,7) Temperature and inner radius of the diskbb model. If temperature error is not given, then temperature is fixed at the value from the period B (denoted with $^*$).
 \label{tab:erosita}}

\end{table*}

}

\section{Discussion}
\label{sect:discussion}

\subsection{AT2019vcb as a QPE}

\begin{figure*}
    \centering
    \includegraphics[width=1\textwidth]{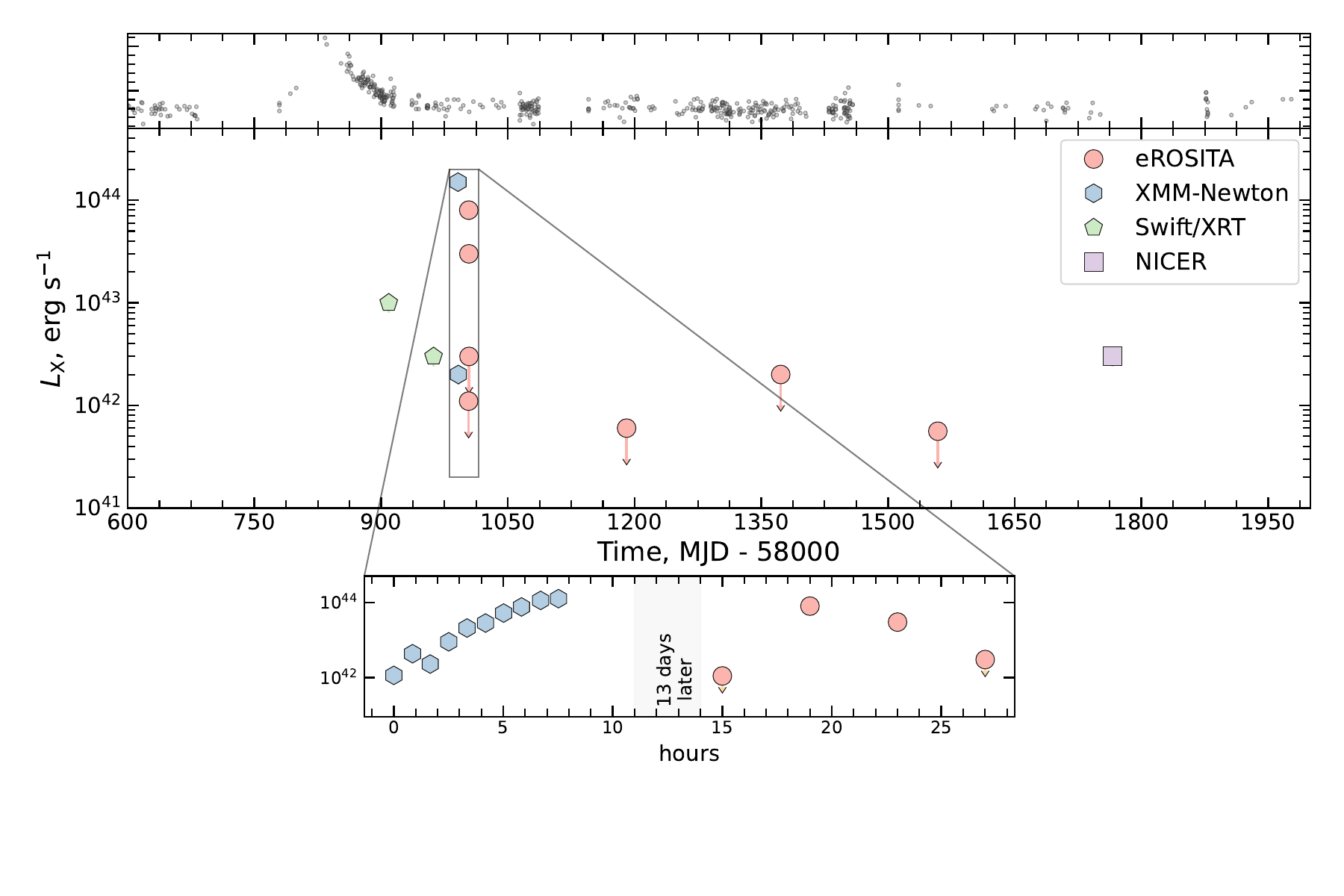}
    \caption{Light curve of AT2019vcb in terms of X-ray luminosity (0.2-2 keV). Various symbols represent luminosity estimated from eROSITA (circles), \textit{XMM-Newton} (hexagons), \textit{Swift/XRT} (pentagon) and \textit{NICER} (square). The inset on the bottom shows the light curves of two assumed QPE flares. The X-ray light curve for \textit{XMM-Newton}, \textit{NICER} and \textit{Swift/XRT} are digitised from the paper by \citeauthor{Quintin2023}.  The top panel shows the optical light curve for guidance. The errors on the light curve are not shown, but they are generally small compared to the amplitude of variation.}
    \label{fig:lc_lumin}
\end{figure*}

\citetalias{Quintin2023} presents strong evidence that the 'half flare' detected by \textit{XMM-Newton} is likely a QPE event. The non-astrophysical sources (e.g. a particle background event) were excluded by the authors through careful data analysis. The alternative astrophysical scenario is that the flare was generated by the TDE itself as a delayed (wrt the optical peak) onset of X-ray emission from the star disruption or a re-brightening flare associated with the existing TDE. Both appear to be unlikely based on the observed properties of the source and prior \textit{Swift/XRT} observation of the TDE decay.  The X-ray luminosity evolution of AT2019vcb is presented in Fig. \ref{fig:lc_lumin}, where some data from \citetalias{Quintin2023} is also shown, including the 'half flare' light curve.

QPE nature of the flare is plausible given its luminosity, spectra and time-scales involved (\citetalias{Quintin2023}). Our findings reported here provide further evidence that AT2019vcb is a bona fide QPE. 
The main observation is that we indeed observe a \textit{short} flare of large amplitude and with a soft spectrum akin to the eight known QPE sources (see introduction). We measure the temperature and emission radius of a black body model as $\sim130\pm10$ eV, $1.6\pm1.0\times 10^6$ km, which are broadly in agreement with the values reported by \citetalias{Quintin2023}, the peak temperature of   $114\pm3$ eV and emission radius in the range of  $2-3\times 10^6$ km.

eRASS1 data provide constraints on the temporal characteristics of the flare.  As QPE events may have rather complicated light curves, definition of their lifetime is not unique. For the purpose of the discussion below we will  operate in terms of full-width at half maximum (FWHM) to determine the duration or rise/fall times of the eruption.  The rise time is likely shorter than 4 hours. If eROSITA missed the peak of luminosity, which is quite likely, the fall time can be somewhat shorter or longer than 4 hours but remains in the several hours range.   The FWHM duration is less than or equal to 8 hours, similar to the flare from \textit{XMM-Newton} data assuming they reached the peak and symmetrical eruption.  However, eROSITA data provide evidence of the asymmetry of the flare similar to eRO-QPE1 \citep{Arcodia2021, Quintin2023}.

If we operate in terms of duration at $\geq$1/10th of the maximum, then the rise time is still less than 4 hours, and fall times are between 4 and 8 hours. The duration in this case is less than 12 hours. Those figures are with the caveat that the quiescent levels were not detected by eROSITA.

 The peak luminosity in eROSITA is $\sim30\%$ smaller than the peak luminosity of the \textit{XMM-Newton} flare, but still large enough to reason that the AT2019vcb has so far the most luminous QPE flares observed.  We also note that the peak luminosity during both QPE events exceeded by at least an order of magnitude the highest luminosity detected by \textit{SWIFT/XRT} from the TDE event itself.

Finally, 13 days passed between the \textit{XMM-Newton} and eROSITA flares, setting the upper limit on the recurrence time of the QPE flares. The lower limit on the recurrence comes from the non-detection of the source during the first three passes in eRASS1, making the lower limit of 8 hours. If we assume that QPE activity was still on during the  eRASS2-4 (6-18 months after eRASS1) then we can increase the lower limit on the QPE recurrence time to 24 hours as no flares were detected during 7-8 passes giving the 24 to 28-hour baseline. However, one should bear in mind that the QPEs could have disappeared by the time of eRASS2-4. Indeed, the reduction in the QPE amplitude over the years was observed in a least one source, eRO-QPE1 \citep{Chakraborty2024, Pasham2024}. The tentative correlation between recurrence time and outburst duration \citep{Arcodia2024b, Chakraborty2024, Nicholl2024} can be used to constrain QPEs in AT2019vcb. If we assume the flare duration of $\sim10-20$ hours, the possible recurrence times are of the order  of $\sim 10-70$ hours \citep[e.g. see fig. 3a in ][]{Nicholl2024}. Note that even if the recurrence time was $\sim 10-20$ hours or so, the non-detection during eRASS 2-4 might simply imply QPE amplitude decay, so even shorter recurrence times are not necessarily in tension with the eROSITA non-detection.

\subsection{AT2019vcb as a TDE}
We discuss the upper limits on the source luminosity derived from eROSITA and the quiescent luminosity from the \textit{XMM-Newton} data in \citetalias{Quintin2023}. Pre-flare upper limit according to the eROSITA data is $1.1\times10^{42}$\ergps, which is slightly below extrapolated TDE afterglow ($L_{\rm X}\propto t^{-5/3}$) ~level of $\sim2\times10^{42}$ \ergps (see fig. 1 from \citetalias{Quintin2023}). One may assume that this discrepancy may arise due to the short-term (weeks) variability of TDE emission afterglow. 

\textit{NICER} data shows the luminosity of $\sim3\times10^{42}$ \ergps~ in July 2022 according to the analysis of \citetalias{Quintin2023}, far above the expected $L_{\rm X}\propto t^{-5/3}$ trend (see their fig. 15). The latest eROSITA sky survey (eRASS4) gives the luminosity upper limit of $\sim5\times10^{41}$ \ergps~ some 7 months earlier (December 2021, see Fig. \ref{fig:lc_lumin}) and is consistent with the $ t^{-5/3}$ law. One may therefore constrain the re-brightening time-scales of the TDE by $\la 7$ months. TDE re-brightening is commonly observed in X-rays \citep{Gezari2017, Malyali2023, Wevers2023}. 

\section{Conclusion}
\label{sect:conclusion}

AT2019vcb is a tidal disruption event (TDE) in a galaxy at redshift $z=0.088$. We reported the  discovery of a short soft X-ray flare from this source with the data of the eROSITA telescope aboard the SRG satellite. The flare occurred on June 3, 2020,  during the first SRG all-sky survey. eROSITA data constrained the flare duration to $\la 12$ hours, its amplitude exceeded a factor of 70 with respect to the quiescent emission level, and its luminosity reached $8\times10^{43}$ \ergps. The spectrum during the flare was thermal, with a temperature of $\sim170$ eV in the accretion disc model.

The flare occurred 13 days after a similar flare was discovered in \textit{XMM-Newton} data by \citet{Quintin2023}. Both events bear very similar temporal and spectral characteristics. Although \textit{XMM-Newton} observed only the rising part of the flare,  \citet{Quintin2023} proposed that this may be a QPE event. eROSITA discovery of the second flare and mapping of its full duration including the return to the quiescence confirms that AT2019vcb/4XMM J123856.3+330957/SRGe J123856.5+330954 is a bona fide QPE source.

eROSITA and \textit{XMM-Newton} data place constraints on the recurrence time of eruptions $\leq13$ days.
Subsequent eROSITA scans show no sign of flaring activity and put a lower limit on the recurrence time of at least 8 hours and 24-28 hours assuming that QPE activity has not ceased 6 months later. Using the relation between QPE duration and recurrence time observed in other sources, we may expect the recurrence time of the order of 60 hours for AT2019vcb.

Our findings provide further evidence that AT2019vcb is a source of QPEs. Should the QPE nature of the source be true, it would strengthen the connection between the population of QPEs and TDEs. It will also expand the diversity of observed QPE characteristics as one with the most luminous flares and the shortest formation time.

\section*{Acknowledgements}

This work is based on observations with the eROSITA telescope onboard the SRG observatory. The SRG observatory was built by Roskosmos in the interests of the Russian Academy of Sciences represented by its Space Research Institute (IKI) in the framework of the Russian Federal Space Program, with the participation of the Deutsches Zentrum für Luft-und Raumfahrt (DLR). The SRG/eROSITA X-ray telescope was built by a consortium of German Institutes led by MPE, and supported by DLR.  The SRG spacecraft was designed, built, launched and is operated by the Lavochkin Association and its subcontractors. The science data are downlinked via the Deep Space Network Antennae in Bear Lakes, Ussurijsk, and Baykonur, funded by Roskosmos. The eROSITA data used in this work were processed using the eSASS software system developed by the German eROSITA consortium and proprietary data reduction and analysis software developed by the Russian eROSITA Consortium. SDB acknowledges partial support by the subsidy FZSM–2023–0015 allocated to Kazan Federal University for assignments in scientific activities.

The authors are grateful to D.R.Pasham and R. Arcodia for inspiring and useful discussions and comments on the manuscript. 

We thank the anonymous referee for useful and constructive comments and suggestions which helped to improve the presentation of our results.

Software: AstroPy \citep{astropy:2018}, Pandas \citep{reback2020pandas},  NumPy \citep{Harris2020}, Matplotlib \citep{Hunter2007}. 
This research has made use of the SIMBAD database, operated at CDS, Strasbourg, France \citep{Wenger2000}. 

\section*{Data Availability Statement}
eROSITA data for this source can be made available upon a reasonable request. Data from ZTF light curves are available via \href{http://db.ztf.snad.space}{SNAD database} \citep{Malanchev2023}.



\bibliographystyle{mnras}
\bibliography{at2019vcb} 



\appendix
\section{eRASS1-4 count rate light curves}

\begin{figure*}
    \centering
    \includegraphics[width=1\textwidth]{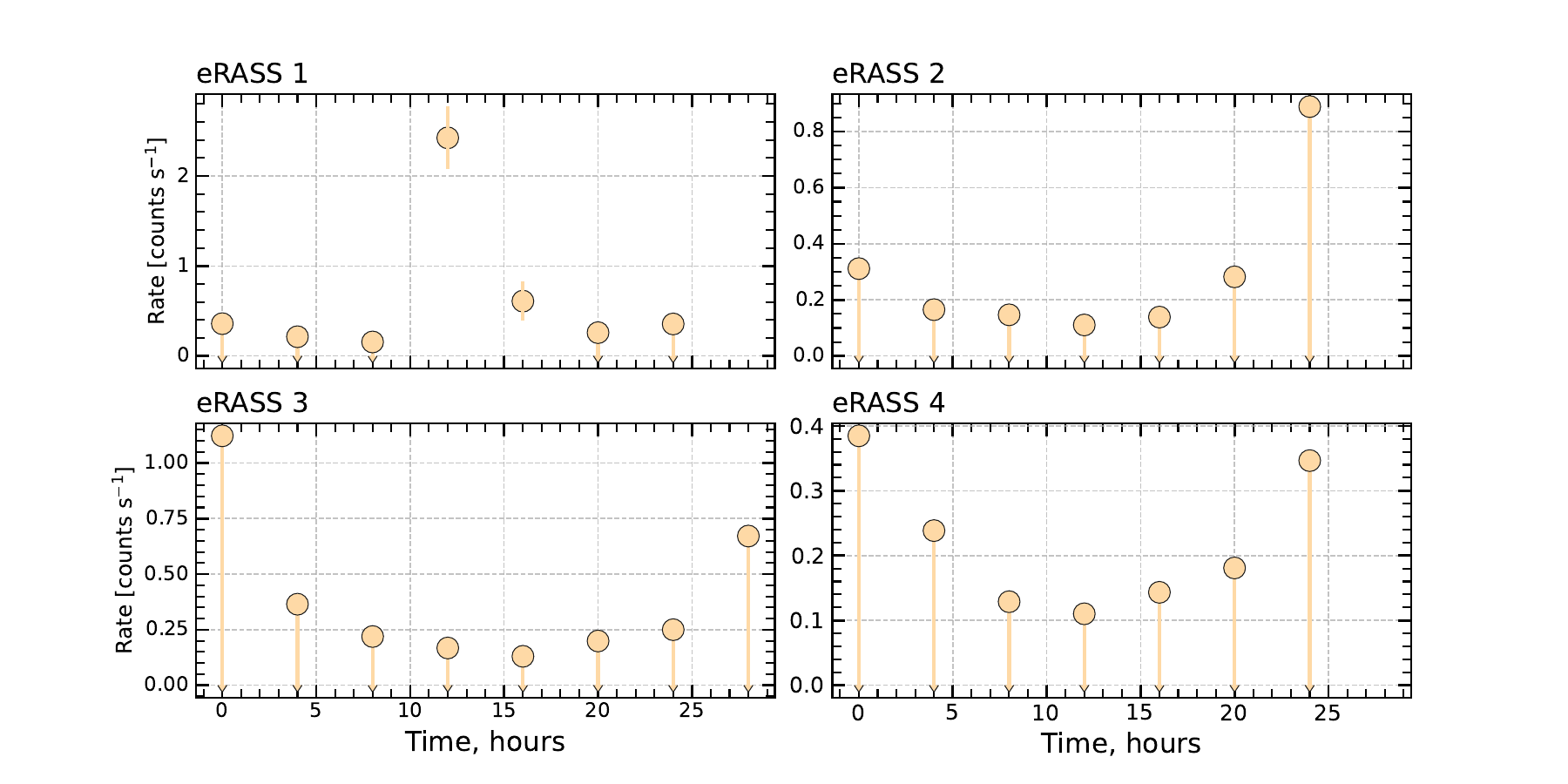}
    \caption{X-ray light curves of AT2019vcb. Each panel corresponds to one eROSITA sky survey, as indicated in the top of each panel. Each point in the light curves corresponds to an individual pass through the source. Upper limits are at 90\% confidence. To convert count rate to luminosity, one may use a conversion factor of $3.25\times10^{43}$ erg evaluated using best fit spectral model in the peak of the flare.}
    \label{fig:lc_all}
\end{figure*}

Fig. \ref{fig:lc_all} shows X-ray light curves of AT2019vcb in individual eROSITA passes through the source. Upper limits are at 90\% confidence and were computed using \citet{Gehrels1986} prescription. The count rate can be converted to the source luminosity using the conversion factor of $3.25\times10^{43}$ erg which was  computed for the best fit spectral model in the peak of the flare.

\section{\texttt{bbodyrad} model}
To ease the comparison with other works, we present results for spectral fits with the \texttt{bbodyrad}-based model (Table. \ref{tab:erosita_bbodyrad}).

{\fontsize{7}{9}\selectfont
\begin{table*}
\centering

\begin{tabular}{lllllll}
\toprule
eRASS$^{(1)}$  &  Period$^{(2)}$  &                        Date$^{(3)}$  &  $F_{\rm X}$, erg s$^{-1}$ cm$^{-2}$ $^{(4)}$  &    $L_{\rm X}$, erg s$^{-1}$$^{(5)}$  &    kT, eV$^{(6)}$  &                      $R$, km$^{(7)}$ \\
\midrule
      eRASS 1  &               A  &             2020-06-03 02:22--10:22  &                         $<5.4 \times 10^{-14}$ &                 $<9.7 \times 10^{41}$ &            $125^*$ &                                    - \\
      eRASS 1  &               B  &                    2020-06-03 14:22  &                   $4.0 \pm0.5 \times 10^{-12}$ & $7.1 \,^{+0.8}_{-0.9} \times 10^{43}$ & $126 \,^{+8}_{-9}$ & $1.6 \,^{+0.8}_{-1.0} \times 10^{6}$ \\
      eRASS 1  &               C  &                    2020-06-03 18:22  &                   $1.6 \pm0.4 \times 10^{-12}$ & $2.8 \,^{+0.7}_{-0.8} \times 10^{43}$ &            $125^*$ & $1.0 \,^{+0.5}_{-0.6} \times 10^{6}$ \\
      eRASS 1  &               D  &  2020-06-03 22:22--2020-06-04 02:22  &                         $<1.6 \times 10^{-13}$ &                 $<2.8 \times 10^{42}$ &            $125^*$ &                                    - \\
      eRASS 2  &               E  &  2020-12-07 01:07--2020-12-08 01:07  &                         $<3.1 \times 10^{-14}$ &                 $<5.6 \times 10^{41}$ &            $125^*$ &                                    - \\
      eRASS 3  &               F  & 2021-06-07 11:22--2021-06-08 15:22   &                         $<7.5 \times 10^{-14}$ &                 $<1.3 \times 10^{42}$ &            $125^*$ &                                    - \\
      eRASS 4  &               G  & 2021-12-10 14:07--2021-12-11 14:07   &                         $<2.7 \times 10^{-14}$ &                 $<4.8 \times 10^{41}$ &            $125^*$ &                                    - \\
\bottomrule

\end{tabular}

 \caption{SRG/eROSITA observations of AT2019vcb. (1) - all-sky sky number; (2, 3) particular observation date;  (4,5) X-ray flux and luminosity in the 0.2--2.0 keV energy range; (6,7) Temperature and estimated radius of the bbodyrad model. If temperature error is not given, then temperature is fixed at the value from the period B (denoted with $^*$).
 \label{tab:erosita_bbodyrad}}

\end{table*}

}


\bsp	
\label{lastpage}
\end{document}